\documentclass[a4paper,preprint,superscriptaddress,preprintnumbers,nofootinbib]{revtex4}

\newcommand{\Slash}[1]{{\ooalign{\hfil/\hfil\crcr$#1$}}} 
\newcommand{\bvec}[1]{\mbox{\boldmath $#1$}}

\usepackage[dvipdf,dvips,dvipdfmx]{graphicx}
\usepackage{amsmath}
\usepackage{amssymb}
\usepackage{float}
\usepackage{wrapfig}
\usepackage{bm}
\usepackage{mediabb}
\begin{document}

\title{The RG flow of Nambu--Jona-Lasinio model at finite temperature and density}

\author{Ken-Ichi \surname{Aoki}}
\email{aoki@hep.s.kanazawa-u.ac.jp}
\affiliation{Institute for Theoretical Physics, Kanazawa University, Kanazawa 920-1192, Japan}

\author{Masatoshi \surname{Yamada}}
\email{masay@hep.s.kanazawa-u.ac.jp}
\affiliation{Institute for Theoretical Physics, Kanazawa University, Kanazawa 920-1192, Japan}

\preprint{KANAZAWA-15-03}

\begin{abstract}
We study the Nambu--Jona-Lasinio model at finite temperature and finite density by using the functional renormalization group. 
The RG flows of the four-Fermi coupling constant in the NJL model are investigated. 
We obtain the chiral phase boundary in cases of the large-$N$ leading approximation and an  improved approximation.
The large-$N$ nonleading term at the vanishing temperature has a singularity at the Fermi surface.
We show that the quantum corrections by the large-$N$ nonleading term largely influence the phase boundary at the low temperature and high density region. 
\end{abstract}
\maketitle

\section{Introduction}
One of the important subjects in elementary particle physics is to understand the basic properties of hadronic matter described by Quantum Chromodynamics (QCD).
The phase diagram and equation of state of QCD have been investigated both theoretically and experimentally.
A number of studies indicate that QCD at finite temperature and finite density has various phases such as the quark-gluon plasma phase with effective chiral symmetry and the color superconductivity phase.

Investigating QCD vacuum need non-perturbative method due to strong interactions of QCD.
Various non-perturbative methods are employed in order to study the phase structure of QCD.
Lattice gauge simulation is very powerful because it is the first principle calculation respecting the gauge symmetry.
However it is difficult to maintain the chiral symmetry on the lattice.
The even more difficult problem which is known as the sign problem occurs at finite chemical potential because the Dirac operator becomes complex. 
Other non-perturbative methods, such as the mean-field approximation (MFA), the Schwinger-Dyson equations (SDE) or the large-$N$ expansion, have been applied to QCD.
These methods do not have the difficulty of maintaining the chiral symmetry nor the sign problem.
However, their approximations break the gauge symmetry and its systematic improvement is not easy.

In this paper, we study the dynamical chiral symmetry breaking (D$\chi$SB) by using the functional renormalization group (FRG), another non-perturbative method~\cite{Aoki:2000wm,Pawlowski:2005xe,Gies:2006wv,Wegner:1972ih,Polchinski:1983gv,Wetterich:1992yh}.
The main idea of the FRG is solving the effective action $\Gamma_\Lambda$ generated by the integration of fluctuations with higher momentum modes $\Lambda < |p| <\Lambda _0$ where $\Lambda$ is the infrared (IR) cut-off scale and the bare action is defined at the initial scale $\Lambda_0$.
The evolution of the effective action with decreasing the IR cut-off scale is described by the functional differential equation, so-called the Wetterich equation~\cite{Wetterich:1992yh,Morris:1993qb},
\begin{align}
\partial_t \Gamma_\Lambda =\frac{1}{2} {\rm STr} 
\left\{ \left[ \frac{\overrightarrow \delta}{\delta \Phi} \Gamma_\Lambda \frac{\overleftarrow \delta}{\delta \Phi} +R_\Lambda \right] ^{-1}\cdot (\partial _tR_\Lambda) \right\}.
\end{align}
This equation itself is exact. However, we cannot solve it exactly.
Therefore we solve the effective action in the truncated theory space.
The improvement of approximation can be done by enlarging the theory space, which are easier and more systematic than the case of the MFA or the SDE.

The D$\chi$SB has been investigated by using low energy effective models such as the Nambu--Jona-Lasinio (NJL) model~\cite{Nambu:1961tp,Nambu:1961fr,Hatsuda:1994pi,Kodama:1999if,Shimizube:2003bv,Braun:2011pp,Braun:2011fw,Braun:2012zq} which is described by the four-Fermi interactions. 
Re-bosonization methods~\cite{Aoki:1999dw,Gies:2001nw,Gies:2002hq} are often applied in order to obtain the physical quantities in macro scale from the interactions in micro scale via the FRG.
In such analyses, however, the theory space is widely expanded and the analysis becomes highly complicated.
Recently, a new method called the ``weak solution'' is proposed in order to evaluate the macro physical values without re-bosonization.
So far, the weak solution method can be applied to a specific type of the approximated FRG equation corresponding to the Fermionic large-$N$ leading approximation.

In this paper we analyze the NJL model at finite temperature and finite density without re-bosonization method nor weak solution method.
Our theory space and quantum corrections included can not be treated by the weak solution method up to the present.
We compare the behaviors of the RG flows of the four-Fermi coupling constant between the large-$N$ leading approximation and that beyond it.
We actually evaluate the RG flows of the inverse four-Fermi coupling constant although its physical and mathematical ground is not completely clear.
In this manner we define macro physics beyond the MFA in the NJL model.
However our method does not allows us to evaluate a critical endpoint in the phase diagram. 
Our work aims to motivate future studies of phase diagrams including the first order phase transition.

This paper is organized as follows: In section \ref{section2} we briefly explain the NJL model in view of FRG and our analysis method.
The RG equations of the four Fermi coupling constant are numerically analyzed at finite temperature and density in section \ref{section3}.
Summary and Discussion are given in section \ref{section5}.
%%
%%%%%%%%%%%%%%
%%%%%%%%%%%%%%
\section{Nambu--Jona-Lasinio model in view of FRG}\label{section2}
In this paper, we study the NJL model as a low energy effective model of QCD. 
The NJL Lagrangian is given by
\begin{align} 
{\mathcal L}_{\rm NJL}={\bar \psi}i{\Slash \partial}\psi 
						+\frac{G}{2N}\{ ({\bar \psi}\psi)^2 + ({\bar \psi}i\gamma_5\psi)^2\} ,
\end{align}
where $G$ is the four-Fermi coupling constant and $N$ is the number of degrees of freedom for the Dirac Fermion ${\bar \psi}$, $\psi$.
The four-Fermi interactions are generated by the QCD gauge interactions.
This Lagrangian is invariant under the chiral U(1) transformation:~$\psi \to e^{i\gamma^5\theta}\psi$.
In present paper, we analyze the ${\rm U}(1)_L \times {\rm U}(1)_R$ symmetric system to study basic structures and behaivors of the four-Fermi interaction at finite temperature and density.
The four-Fermi coupling constant in the effective action corresponds to the chiral fluctuation: $G \sim \langle ({\bar \psi}\psi)^2\rangle $.
Therefore, the divergence of the four-Fermi coupling constant in the course of RG flow means the signal of the D$\chi$SB as the second order phase transition. 
The RG flow equation of the four-Fermi coupling constant for the $4d$ sharp cut-off scheme in the large-$N$ leading approximation is given by
\begin{align}
\partial _t g=-2g+2g^2
\end{align}
where $t$ is the dimensionless RG scale defined by $t=\log (\Lambda_0/\Lambda)$, and $g=G\Lambda^2/4\pi^2$ is the dimensionless four-Fermi coupling constant~\cite{Aoki:1999dv}. 
This RG equation has an ultraviolet fixed point $g^\ast =1$. 
When we solve this equation with an initial value $g_0>1$, the RG flow diverges at the following finite scale,
\begin{align}
t_{\rm c}=\frac{1}{2}\log \left( \frac{g_0}{g_0-1} \right) .
\end{align} 
We cannot solve the RG flow after $t_{\rm c}$ because of this divergence. 
Recently, the ``weak solution'' method~\cite{Aoki:2014ola} is introduced in order to define the flows after this divergence of $g$. 
In this section, we propose another analysis method to evaluate the RG flow of $g$. \\
\indent We define the RG flow equation of the inverse four-Fermi coupling constant, ${\tilde g}=1/g$, and we have the RG flow equation of ${\tilde g}$ as follows,
\begin{align}
\label{rgeqgs}
\partial _t {\tilde g}=2{\tilde g}-2.
\end{align}
The solution of this equation is 
\begin{align}
{\tilde g}(t)=({\tilde g}_0-1)e^{2t}+1.
\end{align} 
The flow with an initial value ${\tilde g}_0<1$ reaches zero at $t_{\rm c}$, ${\tilde g}(t_{\rm c})=0$, which corresponds to the divergence of $g$.
In order to interpret the RG flow with negative ${\tilde g}$ values, the Euclidean NJL action is bosonized,
\begin{align}
S_{\rm NJL}=\int d^4x \left[
{\bar \psi}({\Slash \partial} +\gamma_0\mu)\psi -\frac{h^2}{G}(\sigma ^2+\pi^2)-h{\bar \psi}(\sigma+i\gamma _5\pi)\psi
\right],
\end{align}
where $h$ is the Yukawa coupling constant and, $\sigma$ and $\pi$ are auxiliary fields.
The inverse four-Fermi coupling constant corresponds to the mass of the mesonic potential:
\begin{align}\label{mandg}
m^2=\frac{h^2}{G}
\end{align}
Therefore the negative RG flow of ${\tilde g}$ might indicate that the curvature of the mesonic potential at the origin becomes negative and the potential takes a double-well shape, thus, D$\chi$SB develops.
Such a crude treatment has turned out to be right in case when the weak solution can be defined~\cite{Aoki:2014ola}.  
This simple analysis method follows the RG flow after D$\chi$SB and reach the IR limit $\Lambda \to 0$. 
In the next section, this method is applied to the NJL model at finite temperature and finite density. 

\section{The RG flow at finite temperature and density}\label{section3}
\subsection{RG flow equations of the four-Fermi coupling constant}

%%%%%
We analyze the Euclidean effective action in the local potential approximation~\cite{Hasenfratz:1985dm},
\begin{align}\label{eaa}
\Gamma_\Lambda[\psi,{\bar \psi}]=
\int _0^\beta d \tau
\int d^3x
							\left[{\bar \psi}({\Slash \partial}+\mu)\psi 
						-\frac{G}{2N}\{ ({\bar \psi}\psi)^2 + ({\bar \psi}i\gamma_5\psi)^2\} \right],
\end{align}
where $\beta$ is the inverse temperature $1/T$ and $\mu$ is the chemical potential.
The evolution of $\Gamma_\Lambda$ is described by the Wetterich equation. In this case it reads,
\begin{align}
\partial _t \Gamma_\Lambda[\psi,{\bar \psi}]
=-{\rm Tr}\left[ \frac{\partial _tR_\Lambda}{\Gamma^{(1,1)}_\Lambda+R_\Lambda} \right],
\end{align}
where $\partial _t=-\Lambda \frac{\partial}{\partial \Lambda}$ and $R_\Lambda$ is the cut-off profile function for the momentum of Fermion defined in Appendix~\ref{thresholdapp}.
We introduce the simple notation,
\begin{align}
\Gamma_\Lambda^{(1,1)}=\frac{\overrightarrow \delta}{\delta {\Psi}^{\rm T}(-p)}\Gamma_\Lambda \frac{\overleftarrow \delta}{\delta \Psi(p)},
\end{align}
with $\Psi^{\rm T}(-p):=(\psi^{\rm T}(-p),{\bar \psi}(p))$ and
$$
\Psi(p):=
\begin{pmatrix}
\psi(p)\\
{\bar \psi}^{\rm T}(-p)
\end{pmatrix}.
$$
The ``Tr'' denotes sum over momenta and internal indices.
%%
%%%%%%%%%%%%%%%%
%%%%%%%%%%%%%%%%
\begin{figure}
\begin{center}
\includegraphics[width=150mm]{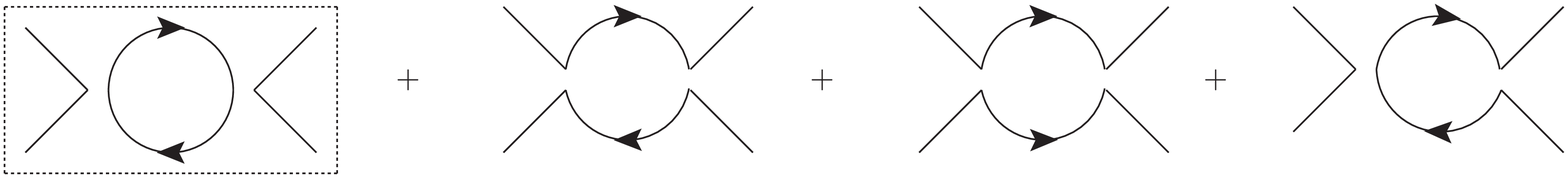}
\end{center}
\caption{The quantum corrections to the four-Fermi interactions. The diagram surrounded by the dashed line is the large-$N$ leading term. The arrow denote the direction of particle number flow.}
\label{fourFermi}
\end{figure}
%%%%%%%%%%%%%%%%
%%%%%%%%%%%%%%%%

Let us briefly explain how to introduce the RG flow equation of the four-Fermi coupling constant.
The modified inverse propagator $\Gamma^{(1,1)}_\Lambda+R_\Lambda$ can be divided into the field-independent part and the field-dependent part,
\begin{align}
\Gamma^{(1,1)}_\Lambda+R_\Lambda ={\mathcal S}_\Lambda+{\mathcal V}_\Lambda[\Psi],
\end{align}
with
\begin{align}
{\mathcal S}_\Lambda :=
\begin{pmatrix}
0	&	-i{\Slash p}^{-}{}^ {\rm T}\\
-i{\Slash p}^{+}	&	0
\end{pmatrix},
\end{align}
and
\begin{align}
{\mathcal V}_\Lambda 
&:=\frac{\overrightarrow \delta}{\delta {\Psi}^{\rm T}}
\left[
-\frac{G}{2N}\{ ({\bar \psi}\psi)^2 + ({\bar \psi}i\gamma_5\psi)^2\} 
\right]
\frac{\overleftarrow \delta}{\delta \Psi} \nonumber \\
&=\frac{G}{N}
\begin{pmatrix}
{\mathcal C}	&	 {\mathcal A}^{\rm T} + {\mathcal B}^{\rm T} \\
{\mathcal A} + {\mathcal B}	&	{\mathcal D}
\end{pmatrix},
\end{align}
where we employ the following notations
\begin{align}
{\Slash p}^{+}& = {\Slash p}+i\mu\gamma_0	+{\bvec {\Slash p}}r_\Lambda({\bvec p})
=\left( p_0 +i\mu ,{\bvec p}(1+r_\Lambda({\bvec p})) \right)_\mu \gamma_\mu, \\
{\Slash p}^{-}& = {\Slash p}-i\mu \gamma_0 + {\bvec {\Slash p}}r_\Lambda({\bvec p})
=\left( p_0 -i\mu ,{\bvec p}(1+r_\Lambda({\bvec p})) \right)_\mu \gamma_\mu,  \nonumber \\
{\mathcal A}&= - \{ \delta_{ij} ({\bar \psi}\psi) + i\gamma^5_{ij}({\bar \psi}i\gamma_5\psi)\} ,\\
{\mathcal B}&= -\{ \psi_i  {\bar \psi}_j + (i\gamma_5\psi)_i ({\bar \psi}i\gamma^5)_j \} , \\
{\mathcal C}&=  \{ \bar \psi_i  {\bar \psi}_j +(\bar \psi i\gamma_5)_i ({\bar \psi}i\gamma^5)_j \} ,\\
{\mathcal D}&=  \{ \psi_i  \psi _j + (i\gamma_5 \psi )_i ( i\gamma^5 \psi)_j \}.
\end{align}
In order to obtain the RG flow equation, we power expand the Wetterich equation with respect to ${\mathcal V}_\Lambda[\Psi]$ as follows,
\begin{align}
\nonumber
\partial_t\Gamma_\Lambda
&=-{\rm Tr}[{\tilde \partial}_t~{\rm ln}({\mathcal S}_\Lambda+{\mathcal V}_\Lambda[\Psi])]\\
\label{modwet}
&=-{\rm Tr}[{\tilde \partial}_t~{\rm ln}({\mathcal S}_\Lambda)]
+{\rm Tr}[{\tilde \partial}_t({\mathcal S}_\Lambda^{-1}{\mathcal V}_\Lambda[\Psi])]
-\frac{1}{2}{\rm Tr}[{\tilde \partial}_t({\mathcal S}_\Lambda^{-1}{\mathcal V}_\Lambda[\Psi])^2]+\cdots ,
\end{align}
where ${\tilde \partial}_t$ acts only on the $\Lambda$ dependence of function $R_\Lambda$.
%%
%The four-Fermi interactions are included in the third term of the second line in Eq.~(\ref{modwet}).
%%
We evaluate the third term as follow:
\begin{align}
-\frac{1}{2}{\rm Tr}[{\tilde \partial}_t({\mathcal S}_\Lambda^{-1}{\mathcal V}_\Lambda[\Psi])^2]
=\frac{1}{2}\int \frac{d^4p}{(2\pi)^4} {\tilde \partial}_t\left[\frac{1}{(p^+)^2} + \frac{1}{(p^-)^2} \right] \frac{G^2}{N^2}
\left({\rm tr}[\gamma_\mu {\mathcal A}\gamma_\mu {\mathcal A}]\right) + \cdots,
\end{align}
where ${\rm tr}[\gamma_\mu {\mathcal A}\gamma_\mu {\mathcal A}] = ({\rm tr}[{\bvec 1}_{\rm spinor}]\times  {\rm tr}[{\bvec 1}_{\rm flavor}]) {\mathcal O}_{\rm 4Fermi}  =4N\{ ({\bar \psi}\psi)^2 + ({\bar \psi}i\gamma_5\psi)^2\} $.
This term yields factor $N$, thus it is the large-$N$ leading term.
Other terms including the operators ${\mathcal B}$, ${\mathcal C}$ and ${\mathcal D}$ do not yield it.
For example, 
\begin{align}\label{example4}
{\rm tr}[\gamma_\mu
 {\mathcal D}\gamma_\mu^{\rm T} {\mathcal C}]
&={\rm tr}[ (\gamma_\mu)_{ij} 
\{ \psi_j \psi _k + (i\gamma_5 \psi )_j ( i\gamma^5 \psi)_k \} 
(\gamma_\mu ^{\rm T})_{kl}
\{ \bar \psi_l  {\bar \psi}_m +(\bar \psi i\gamma_5)_l ({\bar \psi}i\gamma^5)_m \}
] \nonumber\\
&=2\{ ({\bar \psi}\gamma_\mu \psi)^2 - ({\bar \psi}\gamma_5\gamma_\mu\psi)^2 \}.
\end{align}
Using the Fierz transformation, we obtain the opertator $({\bar \psi}\psi)^2 + ({\bar \psi}i\gamma_5\psi)^2$.
Although the nonleading terms yields not only the scalar-type operator but also other four-Fermi operators such as a vector-type operator $({\bar \psi}\gamma_\mu \psi)^2 + ({\bar \psi}\gamma_5\gamma_\mu\psi)^2$, these terms are dropped in our truncated theory space. 
The possible ways of contraction between the operators are exhibited by the Feynman diagrams as shown in Fig.~\ref{fourFermi}.
%The Feynman diagrams of the quantum corrections generated by the four-Fermi interaction are shown in Fig.~\ref{fourFermi}.
%%
%These diagrams exhibit the possible ways of the Wick contractions.
%%
In particular , the first diagram in the dashed box is the large-$N$ leading term.
The large-$N$ leading approximation neglects other diagrams than the first one.
%%
%%%%%%%%%%%%%%%%%%
%%%%%%%%%%%%%%%%%%
\begin{figure}
\begin{center}
\includegraphics[width=80mm]{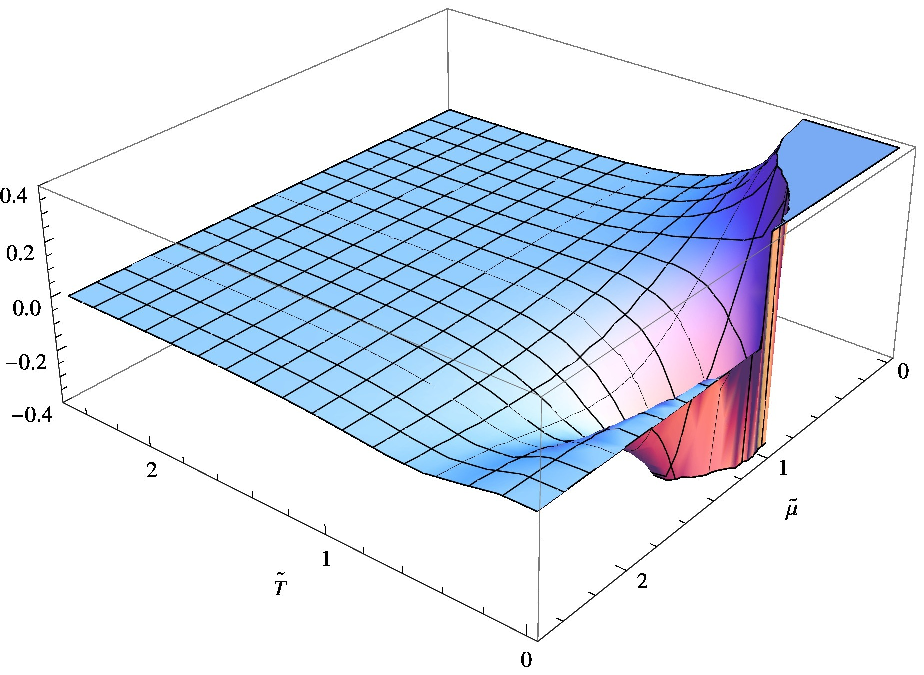}
\includegraphics[width=80mm]{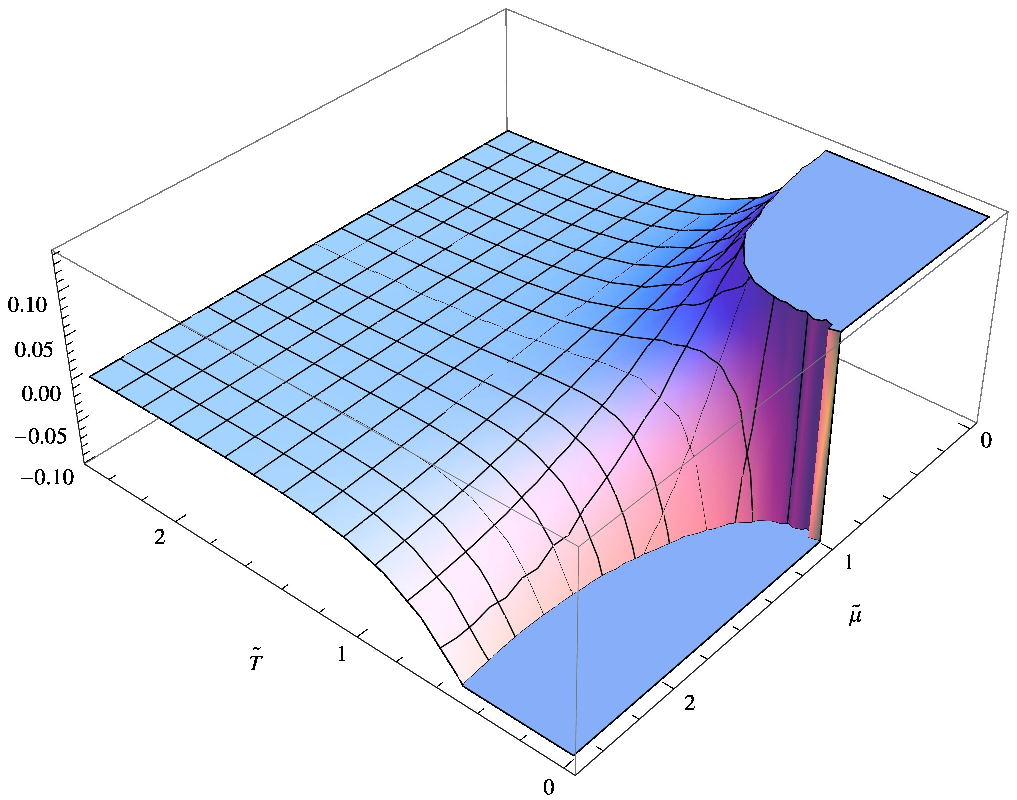}
\end{center}
\caption{The threshold functions $I_0$ (left) and $I_1$ (right) on ${\tilde T}-{\tilde \mu}$ plane. }
\label{threshold}
\end{figure}
%%%%%%%%%%%%%%%%%%
%%%%%%%%%%%%%%%%%

\indent We obtain the RG flow equation of the dimensionless four-Fermi coupling constant, 
\begin{align}\label{njlfinitetmuln}
\partial _t g&=-2g+\frac{1}{3}\{ 4I_0({\tilde T},{\tilde \mu})-\frac{1}{N}I_1({\tilde T},{\tilde \mu})\} g^2,
\end{align}
where the rescaled temperature ${\tilde T}=T/\Lambda$ and rescaled chemical potential ${\tilde \mu}=\mu/\Lambda$, obeys the following equations respectively, 
\begin{align}
\partial _t {\tilde T}&={\tilde T},\\
\partial _t {\tilde \mu}&={\tilde \mu}.
\end{align}
The ``form factor'' $I_0$ and $I_1$ are the threshold functions of ${\tilde T}$ and ${\tilde \mu}$, which are given by the shell mode momentum integration of the large-$N$ leading term and the third diagram in Fig.~\ref{fourFermi}, respectively, and they are defined in Appendix~\ref{thresholdapp}.
The second and fourth diagrams in Fig.~\ref{fourFermi} do not contribute to the RG flow equation of $g$ in the truncated effective action~(\ref{eaa}).
We adopted the $3d$ optimized cut-off function~\cite{Litim:2001up,Litim:2006ag} as the regulator function $R_\Lambda$ which makes it easy to integrate in momentum space with the Matsubara summation~(see Appendix.~\ref{thresholdapp}).
The large-$N$ leading approximated equation can be obtained by taking the limit $N\to \infty$, which makes $I_1$ term vanish. 

Let us now discuss the RG flows at finite temperature and density in a qualitative manner. 
First, we discuss the fixed point structure of the RG equation.
The non-trivial fixed point is given by
\begin{align}
g^\ast=\frac{6}{4I_0({\tilde T},{\tilde \mu})-\displaystyle \frac{1}{N}I_1({\tilde T},{\tilde \mu})},
\end{align}
hence, the fixed point moves with the RG evolutions of dimensionless temperature and density.
When the threshold functions at finite temperature and density are smaller than these with the vanishing temperature and density, we obtain $g^\ast_{T\neq 0, \mu\neq 0}>g^\ast_{T=0,\mu=0}$.
Thus, the chiral symmetry tends to be restored by thermal and finite density effects.

Second, we briefly discuss the effects of the large-$N$ nonleading term at vanishing temperature and density. By taking the limit $T\to 0$ and $\mu\to 0$, we obtain
\begin{align}
\partial_t g=-2g+\frac{1}{3}\left( 4-\frac{1}{N}\right) g^2,
\end{align}
where the factor $4$ in the leading term is caused by the trace of the spinor space. 
Note that the difference of coefficients, $\frac{4}{3}g^2$ here compared with $2g^2$ in Eq.~(\ref{rgeqgs}), comes out of the difference of cut-off schemes.
The nonleading term has negative sign, and it suppresses D$\chi$SB. 

Finally, we discuss finite temperature and density effects.
We show the threshold functions on ${\tilde T}-{\tilde \mu}$ plane in Fig.~\ref{threshold}.
The solutions of the RG equations of ${\tilde T}$ and ${\tilde \mu}$ become simply the exponentially-growing solutions with increasing $t$, thus, the ratio ${\tilde \mu}/{\tilde T}$ is a scale independent constant, $\mu/T$.
Therefore the RG flows of ${\tilde T}$ and ${\tilde \mu}$ move on the straight line with the slope $\mu/T$ on ${\tilde T}-{\tilde \mu}$ plane.
Note that the threshold functions $I_0$ and $I_1$ with the vanishing temperature have the singular point at the Fermi surface $\Lambda=\mu$:
\begin{align}
I_0(0,{\tilde \mu})&=1-\theta({\tilde \mu}-1)-\delta ({\tilde \mu}-1),\\
I_1(0,{\tilde \mu})&=\frac{1}{2(1+{\tilde \mu})^2}+\frac{1}{(1-{\tilde \mu})^2}\left( \frac{1}{2}-\theta({\tilde \mu}-1)\right) +\frac{1}{1-{\tilde \mu}}\delta ({\tilde \mu}-1),
\end{align}
where $\theta({\tilde \mu}-1)$ is the step function.
The function $I_0$ at finite dimensionless chemical potential and the vanishing temperature is constant and positive value.
In particular, this function vanishes in the region $\Lambda \le \mu$.
On the other hand, the function $I_1$ depends on ${\tilde \mu}$ in arbitrary scale. 
Furthermore, this function becomes quite large at the Fermi surface, therefore, the large-$N$ nonleading term is stronger than the leading term.
This means that the RG flow of the four Fermi coupling constant with the large-$N$ nonleading term tends to go to positive infinity at low temperature and high density.
In the next section, we numerically evaluate the RG flow equations.

\subsection{Numerical analysis}
%%%%%%%%%%%%%%%%%%
%%%%%%%%%%%%%%%%%%
\begin{figure}
\begin{center}
\includegraphics[width=160mm]{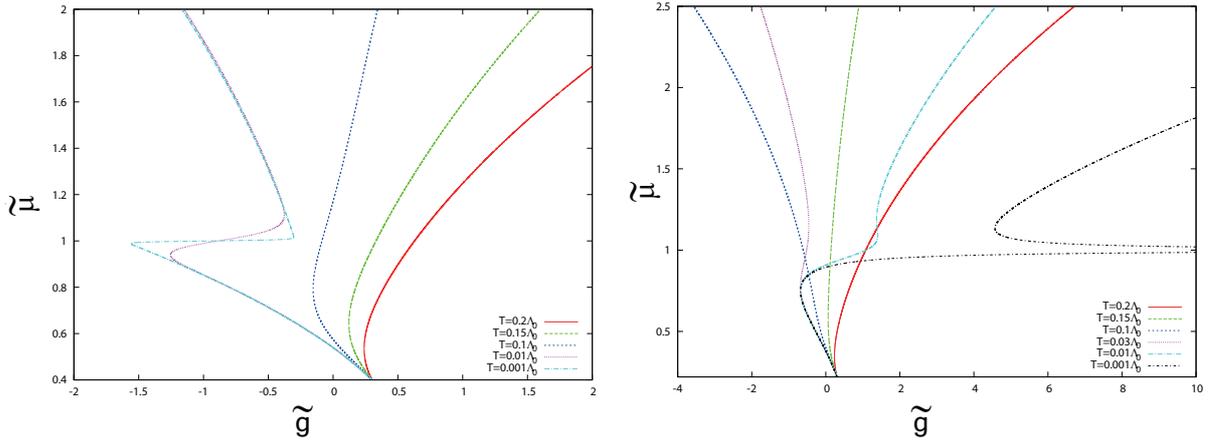}
\end{center}
\caption{The RG flows in case of the large-$N$ leading at ${\tilde \mu}=0.4$ (left) and the large-$N$ nonleading with $N=1$ at ${\tilde \mu}=0.22$ (right). }
\label{flows}
\end{figure}
%%%%%%%%%%%%%%%%%%
%%%%%%%%%%%%%%%%%%
We solve the RG flow equations of the inverse four-Fermi coupling constant ${\tilde g}$. 
The numerical results of the RG flows on ${\tilde g}-{\tilde \mu}$ plane are shown in Fig.~\ref{flows}. 
The left-hand side panel and the right-hand side panel are the large-$N$ leading approximation case and the nonleading extended case, respectively.
Here, we used the initial value: ${\tilde g}=0.3$ where the RG flow at the vanishing temperature and density goes to the negative region in the IR limit. 
First, let us discuss the large-$N$ leading case.
The RG flows of $\tilde g$ at higher temperature cases ($T=0.2 \Lambda_0$ and $T=0.15\Lambda_0$) go to positive infinity. 
The flow at $T=0.1\Lambda_0$ goes to negative region crossing the origin ${\tilde g}=0$, and comes back to positive region. 
Eventually, this flow goes to positive infinity at the IR scale.
The RG flows at lower temperature cases ($T=0.01\Lambda_0$ and $T=0.001\Lambda_0$) go to negative region at the IR limit.

These flows can be evaluated only by the RG equation for the inverse four-Fermi coupling constant.
If we solve the RG equation for the four-Fermi coupling constant Eq.~(\ref{njlfinitetmuln}), these flows stop in the middle of the RG scale.
Solving the RG equations of the inverse four-Fermi coupling constant is useful to discuss the flows after D$\chi$SB, although it is not strictly authorized for large-$N$ nonleading case.
It should be noted that to discuss RG flows towards positive infinity at IR scale do not always imply the restoration of the chiral symmetry.
The RG flow of the inverse four-Fermi coupling constant corresponds to the curvature at origin of the effective potential with respect to the expectation value $\sigma \sim \langle {\bar \psi}\psi \rangle$.
In case of the first order phase transition, the curvature at origin may remain positive, even after the D$\chi$SB. 
We cannot distinguish between the restoration of chiral symmetry and the first order D$\chi$SB.
In other words, we see the effective potential only at the neighborhood of origin.
The chiral phase transition including the first order phase transition should be investigated by searching the global minima of the effective potential by using the weak solution method or the re-bosonization method. 
Although our method can be applied only to the second order phase transition, our results may be used as references for advanced analysis of the chiral phase transition.

Second, let us discuss the nonleading extended case. 
The RG flows at low temperature~($T=0.01\Lambda_0$ and $T=0.001\Lambda_0$) go to positive infinity at the IR limit.
These behaviors are quite different from the large-$N$ leading case.
The large-$N$ nonleading term becomes larger than the leading term because the threshold function $I_1$ has the singularity at $\mu=\Lambda$.
%%

%%%%%%%%%%%%%%%%%%
%%%%%%%%%%%%%%%%%%
\begin{figure}
\begin{center}
\includegraphics[width=130mm]{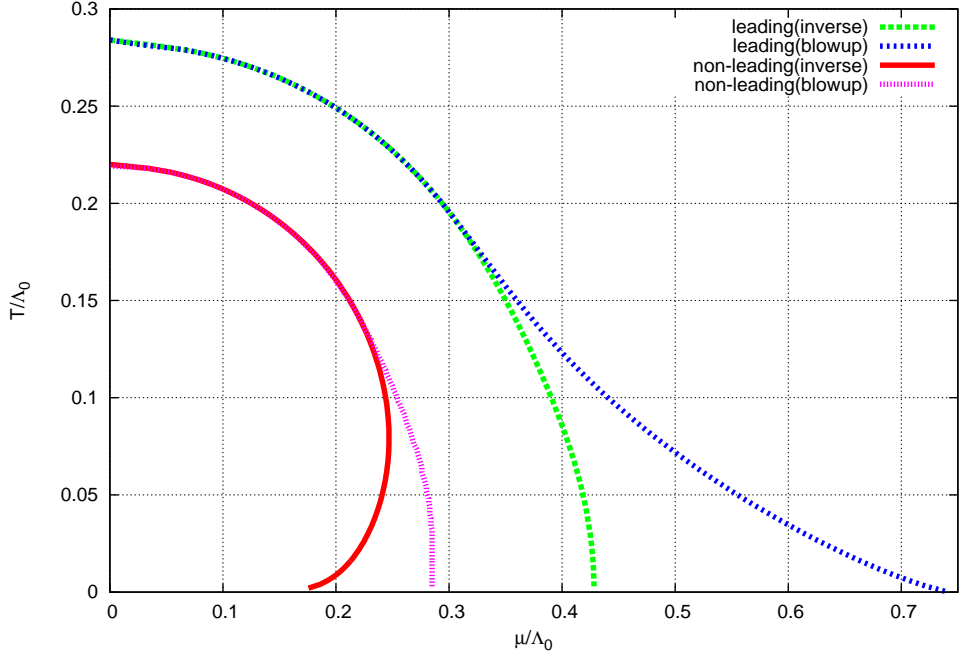}
%\put(-310,100){\Large{$\tilde g <0$}}
%\put(-140,150){\Large{$\tilde g >0$}}
\end{center}
\caption{The chiral phase structure given by the large-$N$ leading case (blue and green line) and nonleading case (red and pink line). We used ${\tilde g}_0=0.3$ as the initial value and $N=1$ for the nonleading case. 
The phase boundaries assume the second order phase transition.}
\label{phase}
\end{figure}
%%%%%%%%%%%%%%%%%%
%%%%%%%%%%%%%%%%%%
The chiral phase diagram on ${\mu/\Lambda_0}-T/\Lambda_0$ plane is shown in Fig.~\ref{phase}.
We show the phase boundaries evaluated by the blowup solution of Eq.~\eqref{njlfinitetmuln} and the RG equation of the inverse four-Fermi coupling constant.
Here we call the solution of the RG equation for the inverse four-Fermi coupling constant simply the inverse solution.

First, we discuss the difference of phase boundaries between the blowup solution case and the inverse solution case.
Since the blowup solution stops the RG flow in the midst of scale, we cannot evaluate it after the chiral symmetry breaking.
Then we notice that the phase boundary with the inverse solution is correct.
In high temeperature and low density region, both boundaries become same.
By contrast, in low temperature and high density region, the phase boundaris evaluated by the blowup solutions become larger than the inverse solutions.
That is, it is especially important to evaluate the RG equation of the inverse four-Fermi coupling constant in this region.

Next, we discuss the difference of phase boundaries with the inverse solution between the large-$N$ leading case and nonleading case.
The D$\chi$SB region becomes smaller in case of the large-$N$ nonleading. 
We find drastic difference of phase boundaries at low temperature and high density region since the large-$N$ nonleading effect strongly suppresses the D$\chi$SB as we previously mentioned.
The chiral susceptibility, the curvature at the origin of mesonic potential $m^2\sim 1/\langle ({\bar \psi}\psi)^2 \rangle$, is sensitive to the nonleading effects at low temperature and high density.

Note that the phase boundaries shown in Fig.~\ref{phase} denote second order phase transition. 
As mentioned below Eq.~\eqref{mandg}, the inverse four-Fermi coupling constant corresponds to the curvature of the effective potential at the origin and its RG equation describes the change of the curvature.
In the present work, the phase boundary is determined by the sign of the inverse four-Fermi coupling constant in IR limit.
The broken phase potential with double well form has the negative curvature of the effective potential at the origin.
%%%
The symmetric phase has a single well form and the positive curvature of the effective potential at the origin.
Thus, the change from the negative curvature to the positive curvature mean the second order phase transition at least near the origin.
One of important statements in our study is that these pictures can be described by the RG flow of the inverse four-Fermi coupling constant.

On the other hand, when the first order phase transition occurs, there might be no global minimum at the origin of the effective potential even in case that the curvature of the effective potential at the origin changes to be positive.
Therefore, it does not always mean we evaluate the global minimum of the effective potential in the present work.
That is, there is possibility of occurring the first order phase transition at larger density region than the second order boundary evaluated in present paper.
%%
%Although the present study does not evaluate the first order phase transition, the behaviors of the curvature the effective potential at the origin include important information of the chiral phase transition and give us the appropriate directions of studies to analyze the system in detail.
%%

%%%%%%%%%%%%%%%%%
%%%%%%%%%%%%%%%%%
%%%%%%%%%%%%%%%%%
\section{summary and discussion}\label{section5}
\begin{figure}
\begin{center}
\includegraphics[width=80mm]{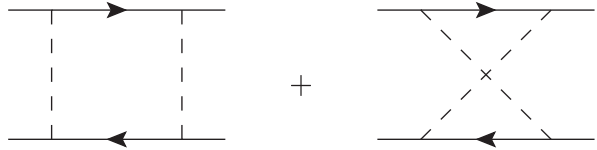}
\end{center}
\caption{The quantum corrections to the four-Fermi interactions generated by the Yukawa interaction. The crossed ladder diagram (right) at the zero temperature and finite density has the singularity at the Fermi surface.}
\label{yukawa}
\end{figure}
In this work, we have studied the RG flows of the four-Fermi coupling constant in NJL model at finite temperature and density by using the FRG. 
We solved the RG equations of the inverse four-Fermi coupling constant.
This manner allows to integrate the equation up to the IR scale without the difficulties of the divergence.
We have investigated the impact of the large-$N$ nonleading term on the RG flow equation. 
The large-$N$ nonleading term has the singularity at the Fermi surface scale $\Lambda=\mu$ with the vanishing temperature, i.e. the quantum fluctuation due to the large-$N$ nonleading term becomes quite large, which makes the system more symmetric, as seen in Fig.~\ref{phase} 

In this analysis, the second order phase transition is investigated. 
In order to investigate the first order phase transition with the critical end point in the NJL model, the global minima of the effective potential with respect to the chiral order parameter must be studied by using more technical methods such as the re-bosonization method or the weak solution method.
Currently, since the weak solution method can be applied to only the Fermionic large-$N$ approximated case,  we need to employ the re-bosonization method.
For example, a truncated effective action of the NJL model with bosonization is given by
\begin{align}\nonumber
\Gamma _{\Lambda}[\psi, {\bar \psi}, \phi]=\int^{1/T}_0d\tau \int d^3x 
& \left\{ Z_{\psi,\Lambda}{\bar \psi}({\Slash \partial} + \mu \gamma _0)\psi 
+{\bar h_{\Lambda}}{\bar \psi}(\sigma +i{\vec \tau}\cdot {\vec \pi}\gamma _5)\psi 
\right. \\
\label{bosoef}
&+\left. \frac{G_\Lambda}{2}[({\bar \psi}\psi)^2+({\bar \psi}i\tau^i\gamma_5 \psi)^2]
+\frac{Z_{\phi,\Lambda}}{2}(\partial _{\mu}\phi^i)^2
+U_{\Lambda}(\phi^2)\right\},
\end{align}
where $\phi^i=(\sigma,{\bvec \pi})$.
The four-Fermi interaction is generated by the Yukawa interaction in this action through the diagrams shown in Fig.~\ref{yukawa}.
The contribution from the crossed ladder diagram at finite density has the singularity at the Fermi surface.
The impact of this singularity to the chiral phase diagram on $T-\mu$ plane should be investigated.
The paper~\cite{Braun:2014ata,Mitter:2014wpa} which study QCD at vanishing temperature and density by using re-bosonization method indicates that the Yukawa coupling constant always gets the same value in the IR scale, i.e. the Yukawa coupling constant at the IR scale does not depend on the initial values. However, the impact of the singularity to the phase boundary is non-trivial.
The analysis of the effective action (\ref{bosoef}) will be presented elsewhere.

On the other hand, the weak solution method is also powerful to analyze the dynamical chiral symmetry breaking~\cite{Aoki:2014ola}.
This method is just developing to apply to many systems.
Therefore, the present work becomes one of valuable benchmarks.
We are going to clearly discuss the relation between our results obtained in this paper and the phase diagram given by the weak solution method in future works.

\section*{Acknowledgements}
We thank Kenji Fukushima, Akira Onishi, Jan. M. Pawlowski, Fabian Rannecke, Daisuke Sato and Motoi Tachibana for discussions and valuable comments.
M. Y. thanks the Yukawa Institute for Theoretical physics, Kyoto University.
Discussions during the YITP workshop YITP-T-13-05 on ``New Frontiers in QCD'' were useful to complete this work. 
M. Y. is supported by a Grant-in-Aid for JSPS Fellows (No.~25-5332). 
K-I. A. is supported by JSPS Grant-in-Aid for Challenging Exploratory Research (No.~25610103).

\begin{appendix}
\section{Threshold functions}\label{thresholdapp}
In this appendix, we show the threshold functions which are obtained by the shell mode integration.
We first introduce the cut-off profile function $R_\Lambda(p)$.
In case of finite temperature and density, we employ the $3d$ optimized cut-off function~\cite{Litim:2001up,Litim:2006ag},
\begin{align}\nonumber
R_\Lambda({\bvec p})
	&=i{\Slash {\bvec p}}\left( \frac{\Lambda}{|{\bvec p}|} -1\right) \theta (1-|{\bvec p}|/\Lambda)\\
	&=:i{\Slash {\bvec p}}~r_\Lambda({\bvec p}),
\end{align}
where $\theta (1-|{\bvec p}|/\Lambda)$ is the Heaviside step function,
\begin{align}
\theta (1-|{\bvec p}|/\Lambda)
=
\begin{cases}
1	&	(|{\bvec p}|<\Lambda),\\
0	&	(|{\bvec p}|>\Lambda).
\end{cases}
\end{align}

We define the following threshold functions,
\begin{align}
{\hat I}_0(T,\mu;\Lambda) 
&=T\sum ^{\infty}_{n=-\infty}\int \frac{d^3p}{(2\pi)^3}{\tilde \partial}_t
\frac{1}{[(p_0+i\mu)^2+{\bvec p}^2(1+r_\Lambda({\bvec p}))^2]},\\
{\hat I}_1(T,\mu;\Lambda) 
&=T\sum ^{\infty}_{n=-\infty}\int \frac{d^3p}{(2\pi)^3}{\tilde \partial}_t
\frac{(p_0+i\mu)(p_0-i\mu)+{{\bvec p}}^2(1+r_\Lambda({\bvec p}))^2}{[(p_0+i\mu)^2+{{\bvec p}}^2(1+r_\Lambda({\bvec p}))^2][(p_0-i\mu)^2+{{\bvec p}}^2(1+r_\Lambda({\bvec p}))^2]},
\end{align}
where ${\tilde \partial}_t=\partial _tr_\Lambda\cdot \partial _{r} $ and $p_0=(2n+1)\pi T$.
We can analytically calculate the above integration and summation and we obtain
\begin{align}
\nonumber
{\hat I}_0(T,\mu;\Lambda) 
&=\frac{\Lambda^2}{3} 
\left. \left[ \left( \frac{1}{2}-n_+\right) + \left( \frac{1}{2}-n_-\right)	+\frac{\partial}{\partial \omega}[n_++n_-]\right]\right| _{\omega\to 1}\\
&=\frac{\Lambda^2}{3} I_0({\tilde T},{\tilde \mu}),\\
\nonumber
{\hat I}_1(T,\mu;\Lambda) 
&=\frac{\Lambda^2}{3}
\left. \left[ \frac{1}{(1+{\tilde \mu})^2} \left( \frac{1}{2}-n_+\right) + \frac{1}{(1-{\tilde \mu})^2}\left( \frac{1}{2}-n_-\right)	+\frac{1}{1+{\tilde \mu}}\frac{\partial}{\partial \omega}n_+   +\frac{1}{1-{\tilde \mu}}\frac{\partial}{\partial \omega}n_-\right]\right| _{\omega\to 1}\\
&=\frac{\Lambda^2}{3} I_1({\tilde T},{\tilde \mu}).
\end{align}
Here we have used the formula,
\begin{align}
\sum^{\infty}_{n=-\infty}\frac{1}{(2n+1)^2\pi^2+\beta^2E_\pm^2}
=\frac{1}{\beta E_\pm}\left( \frac{1}{2}-n_\pm\right),
\end{align}
where $n_\pm$ is the Fermi-Dirac distribution function:
\begin{align}
n_\pm=\frac{1}{\exp(\beta E_\pm)+1}
\end{align}
with $E_\pm=\Lambda\pm \mu$ and $\beta =1/T$.

\end{appendix}

\bibliography{refs}
\end{document}